\documentstyle[12pt,epsf]{article}
\def\reference{\parskip 0pt\par\noindent\hangindent 0.5 truecm}

\begin{document}
%
\title{Big Bang Nucleosynthesis, Population III, and Stellar Genetics in the 
Galactic Halo
\footnote{Invited Talk, Annual Conference of the Astronomical Society of 
Australia, Lorne, 2-4 July 2001
\hfill
Submitted to Publ. Astron. Soc. Australia, 10 September 2001;
in final form 29 October 2001.
\hfill
Accepted 2 November 2001}
}
%


\author{Sean G. Ryan $^{}$ 
} 

\date{}
\maketitle

{\center
$^{}$ Dept of Physics and Astronomy, The Open University, Walton Hall, 
Milton Keynes MK7 6AA, United Kingdom\\
s.g.ryan@open.ac.uk\\[3mm]
}

\begin{abstract}
The diverse isotopic and elemental signatures produced in different 
nucleosynthetic sites are passed on to successive generations of stars.
By tracing these chemical signatures back through the stellar populations of
the Galaxy, it is possible to unravel its nucleosynthetic history  and even to
study stars which are now extinct. 
This review considers recent applications of "stellar genetics" 
to examine the earliest episodes of nucleosynthesis in the Universe, in
Population III stars and the Big Bang.
\end{abstract}


\bigskip

%
%

\section{Stellar genetics}

Nucleosynthesis in the first few minutes following the Big Bang produced 
five isotopes in significant quantities: $^1$H, $^2$H, $^3$He, $^4$He, and
$^7$Li. The last of these accounts for less than one part per billion. 
All other elements were produced later, either in stars or as a consequence of
stellar evolution. 

Stellar nucleosynthesis theory predicts the yields of a wide range of isotopes,
but in most astronomical spectra only elemental (rather than
isotopic) abundances can be measured because the wavelength shift between
different isotopes of an element is usually well below the intrinsic line width 
set by thermal and pressure broadening. Amongst the few exceptions are $^1$H and 
$^2$H (e.g. Linsky et al. 1995; Burles, Kirkman, \& Tytler 1999), 
and $^6$Li and $^7$Li (Smith, Lambert \& Nissen 1993), whose isotope shifts are
particularly large because of their small atomic masses. 
Isotope shifts are usually larger for
molecular features, and $^{12}$C and $^{13}$C (e.g. Brown \& Wallerstein 1989)
and certain Mg isotopes (e.g. Shetrone 1996) can be
distinguished in CH and MgH bands. Although isotope shifts become even smaller
for higher atomic masses, differences in the hyperfine structure of odd and even
isotopes {\it may} allow constraints to be placed on isotope ratios for some
elements as heavy as Ba (Magain 1995) and Eu (Hauge 1972).
Nevertheless, stellar spectral measurements are generally limited to elemental rather 
than isotopic abundances. 

\begin{figure}[ht]
\begin{center}
\leavevmode
\epsfxsize=155mm
\epsfbox[22 440 565 690]{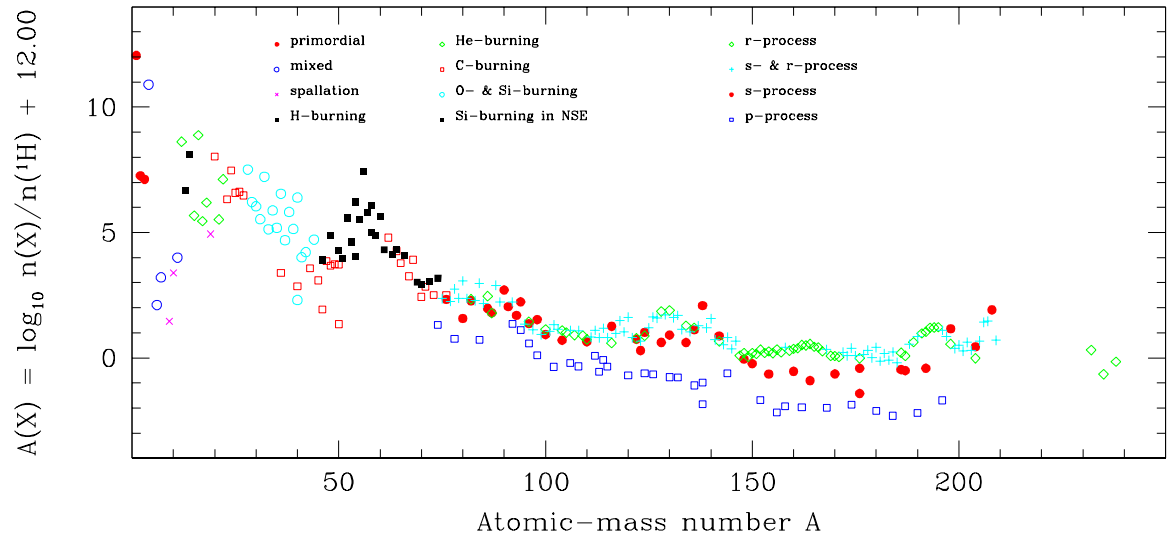}
\end{center}
\caption{Isotopic abundances arranged by atomic-mass number A.
The legend indicates the main nucleosynthesis pathway for each species.
(Based on data from Lang 1980, Table 38).}
\label{sad}
\end{figure}

By measuring the time-evolution of abundances 
coming from a variety
of nucleosynthesis sites --- see Figure~\ref{sad}
--- such as AGB stars, the cores of supernovae, and
various radial zones in the envelopes of massive stars, 
it is possible to discover the roles of a wide range
of stars over the history of the Galaxy. 
This is the activity I have branded ``stellar genetics'': using the 
isotopic and elemental signatures that are passed from one stellar generation
to the next to trace the 
particular nucleosynthesis reactions that have occurred in the past, 
and hence also to trace the sites of those reactions even
when these may no longer exist. For example, low-metallicity asymptotic giant 
branch (AGB) stars have all now become white 
dwarfs, but it is nevertheless possible to study the internal structure of those
AGB stars via the material they have dumped onto the surface of longer-lived
companions (Ryan et al. 2001a).

The supernovae of massive stars are important 
early contributors to the enrichment of the Galaxy. Stars whose main-sequence mass
exceeds 8-10~M$_\odot$ are expected to become core collapse (type II) supernovae
on timescales $<50$~Myr, as short as $\sim$1~Myr for stars of a few tens of solar 
masses.
SN~II produce of order ten times more oxygen than iron (by mass), whereas 
SN~Ia, with lower-mass progenitors and hence longer time lags, 
produce of order seven times more iron than oxygen 
(e.g. Timmes, Woosley \& Weaver 1995).
Consequently, the abundance ratios of oxygen and the
$\alpha$-elements\footnote{$\alpha$-elements are those relatively abundant light
elements whose nuclei may be viewed as multiples of He nuclei, resulting either
from He-burning or further fusion of He-burning products.}
to iron are higher in Galactic halo stars than in the Sun
(e.g. Tinsley 1979, Norris, Ryan \& Beers 2001).

Although supernovae have been widely modelled, the role of stellar winds
in removing processed material from highly luminous stars has not yet been
widely investigated for the early enrichment of the halo, presumably because
of uncertainties in the treatment of mass loss and its lower efficiency in
metal-poor environments. However, 
the unexplained high levels of carbon enrichment in the early Galaxy,
where perhaps 10\% of stars have carbon-to-iron ratios up to 100 times
solar (Norris, Ryan \& Beers 1997a; Barbuy et al. 1997; Rossi, Beers \& Sneden 1999)
suggest that more attention should be devoted
to investigating this means of enriching the early interstellar medium.
This is not to say that stellar winds are the only way of effecting large
carbon excesses; indeed, the carbon-rich stars show a {\it range} of other
signatures including s-process enhancements (Norris et al. 1997a), r-process
enhancements (Cowan et al. 1995), and normal neutron-capture-element abundances
(Norris, Ryan \& Beers 1997c, Aoki et al 2002).
Nevertheless, one may speculate that mass-loss may help produce high 
carbon-to-iron ratios without producing neutron-capture elements 
if the ratio of black-hole
remnants to neutron star remnants is greater in metal-poor than metal-rich populations,
as it might be since metal-poor stars are more compact and hence sit deeper in their gravitational
potential wells. This might manifest itself in a lowering of the progenitor mass which 
corresponds to the 
division between (low-mass) neutron-star remnants and (high-mass)
black-hole remnants in metal poor populations.
(This uncertain value is sometimes set in the range 30-50~M$_\odot$ in Galactic chemical
evolution calculations; see also Prantzos (1994).)
This would allow stellar winds to enrich the interstellar medium
in carbon, yet avoid an associated enrichment in iron due to fall back
onto a neutron star.

Stars below the mass limit for SN~II become AGB stars which are responsible for
most Galactic nucleosynthesis of carbon and s-process elements, but because of their lower 
masses they enrich the Galaxy {\it after} the first SN~II. Models of 8--10~M$_\odot$
supernovae as the source of r-process nuclei suggest that these appeared as
the Galaxy reached [Fe/H] $\simeq$ $-3.0$ (e.g. Mathews \& Cowan 1990),
which implies that AGB stars
will enrich the halo only once this metallicity has been reached. This makes
AGB stars unsuitable as the source of the carbon excess in lower metallicity 
stars.
Also, the s-process contribution is insignificant until an even higher 
metallicity is reached, because of the need for pre-existing seed nuclei 
(Spite \& Spite 1978; Truran 1981). 
The numerical model of Travaglio et al. (1999) suggests that the s-process
appears only at [Fe/H] $\simeq$ $-1$, though
Pagel \& Tautvai\v{s}ien\.{e} (1997) argue for its appearance
as low as [Fe/H] = $-2.5$.

Although the science of stellar spectroscopy
is more than a century old, many of the oldest stars in the Galactic halo have
been discovered only in the last decade, as a result of the 
Beers, Preston \& Shectman (1985, 1992) survey and its successors (e.g. Norris, Ryan
\& Beers 1998; Allende Prieto et al. 2000). 
Furthermore, detailed spectroscopic study of such stars requires
\'echelle spectrographs on 4--10~m-class telescopes equipped with efficient
electronic detectors.
The remainder of this review highlights a number of issues concerning the 
very earliest stages of chemical evolution, namely Big Bang nucleosynthesis and
the lack of identifiable Population III stars.
Recent reviews of the evolution of other elements in the early phases of the 
Galaxy can be found in Ryan (2001) and Ryan et al. (2001b).

\section{The elusive Population III}

The term ``Population III'' means different things to different people. For some 
it refers to very metal-poor stars, whereupon the distinction between Pop~III and
Pop~II becomes blurred.
More usefully it can be applied to the first generation of stars that 
formed from material enriched by Big Bang nucleosynthesis
alone, prior to stellar nucleosynthesis occurring. This is the definition 
that will be adopted here. The massive stars of this first stellar generation 
were probably responsible for the re-ionization of the universe, possibly at 
redshifts around $z \simeq 10$ (Tumlinson \& Shull 2000),
before the formation of Galactic halo stars
at a redshift $z \simeq 5$ (e.g. Edmunds \& Phillipps 1997).

It might be supposed that searches for extremely metal-poor stars should encounter
some having Population III composition, but none has been recognised. In fact, 
few would ever have existed. Recall that
a single 25~M$_\odot$ supernova is sufficient to enrich a $10^6$~M$_\odot$ primordial
gas cloud to [Fe/H] $\simeq$ $-3.5$ (Ryan, Norris \& Bessell 1991), demonstrating
that very little stellar evolution and nucleosynthesis is required to significantly
enrich the interstellar medium. Furthermore, 
such supernovae will appear on timescales
of only a few million years. Pop~III stars will be a very minor constituent of the
Galaxy.

A more instructive question to ask is whether there is a lower limit on the 
metallicity of Pop~II stars. The answer appears to be ``yes''. Despite the 
systematic surveys cited above, no star more metal-deficient than CD$-38^\circ$~245
(Bessell \& Norris 1984) has been discovered, though a few of comparable metallicity
have been found. Although a simple closed-box model for galactic chemical evolution
is a poor model for the Galactic disk, it does a surprisingly good job
of fitting the Galactic halo (Hartwick 1976; Ryan \& Norris 1991a) and predicts
a factor of ten fewer stars for each factor of ten lower metallicity. However,
the good fit breaks down once [Fe/H] falls to $-4$, and a substantial deficit of
stars becomes apparent. Based on current halo samples, the simple model
suggests that 11 stars with [Fe/H] $<$ $-4$ should have been detected, whereas
at most two are known (Norris 1999). The Poisson probability of this occurring by 
chance is just 0.12\%, strongly suggesting the deficit is real.

A number of potential explanations for the absence of the more metal poor stars can be proposed:
\newline $\bullet$ The lowest metallicity (Pop~III?) stars probably
formed before the
Galaxy had collapsed to the size of the halo, and hence the missing stars may have
a greater radial scale than the halo.
\newline $\bullet$ The initial mass function (IMF) may have had a lower limit above 
0.8~M$_\odot$, so no main-sequence or red giant branch Pop~III stars remain after
$\sim$15~Gyr (e.g. Nakamura \& Umemura 1999).
\newline $\bullet$ Since the pre-main-sequence contraction time of 0.8~M$_\odot$
stars is $\sim 10^7$~yr whereas the first supernovae would have exploded after
only $\sim 10^6$~yr, some contamination of low-mass Pop~III stars may have
occurred. However, since low-mass stars will have already contracted to the 
base of their Hayashi tracks and undergone deuterium burning on a $\sim 10^6$~yr
timescale, it is less likely that contamination from a supernova would be
very effective.
\newline $\bullet$ Pop~III stars may have accreted sufficient material on passages
through the Galactic disk that their thin outer convective zones are now
contaminated by heavier elements (e.g. Yoshii 1981). However, such accretion could
not hide Pop~III giants, whose greater convection would give them lower 
metallicities than dwarfs (which is not observed), and their heavy element 
abundance ratios would resemble those of the accreted material, but halo abundance
ratios are different from that of disk material (Ryan \& Norris 1991b).

So, Pop~III stars are undetected, but that does not mean we cannot study them.
Thanks to stellar genetics, the yields of the first Pop~III supernovae
are observable as the post-primordial composition of Pop~II stars. Comparisons
of the abundances of extremely-metal-poor halo stars with the yields predicted
for Pop~III supernovae (e.g. Woosley \& Weaver 1995) search
for concordance in the abundance ratios, 
with the aim of revealing the
mass, supernova-energy ranges, and possibly eventually the rotation of
Pop III stars, thus identifying which types of stars have contributed 
to the early chemical
enrichment of the Galaxy (e.g. Norris et al. 2001).

\section{Primordial isotopes}
\subsection{$^4$He}

Most Pop~II stars are not hot enough to excite He spectral lines,
and those that are show the effects of diffusion which, in the 
absence of convective mixing, causes elements to stratify in the stellar atmosphere.
To find the primordial $^4$He composition, it is necessary 
to measure H II regions in galaxies that are 
less chemically enriched than the Milky Way. 
By measuring the strengths of emission lines of hydrogen, helium, 
and various metals (especially oxygen and nitrogen), 
it is possible to derive relative 
abundances (e.g. Pagel et al. 1992).

The evolution of the helium mass fraction $Y$ with metallicity mass fraction $Z$ 
is revealed in a graph of $Y$ vs $Z$, which
gives a straight line graph of the  form 
$Y = Y_{\rm p} + ($d$Y/$d$Z) Z$.
The primordial helium abundance $Y_{\rm p}$ is simply 
the intercept. 

Challenges in deriving the primordial helium mass fraction $Y_{\rm p}$ include:
\newline 
$\bullet$ Helium production does not always accompany metal production at 
the same rate. That is, d$Y$/d$Z$ is not the same for all objects.
Wolf-Rayet stars enrich H~II regions more in He than the 
metals suggest, and bias $Y_{\rm p}$ upwards. 
Fortunately, Wolf-Rayet stars have a spectral signature 
which allows the most troublesome cases to be detected (e.g Pagel et al. 1992),
but some bias may remain at a low level.
\newline 
$\bullet$ H II regions are visible only because the gas is excited by the 
UV thermal radiation from hot stars embedded in the clouds. 
The absorption lines in the stellar spectra literally undermine 
the emission line intensities,
resulting in artificially 
weak He emission lines, from which low values of Y$_p$ would be derived
(e.g. Olive \& Skillman 2001).
\newline
$\bullet$ It is often assumed that only UV thermal radiation from the hot stars 
ionises and excites the nebula 
gas, but there may also be some excitation by collisions between gas atoms or by
re-absorption of the emitted radiation. The effect of ionisation correction 
factors is unclear (Ballantyne, Ferland \& Martin 2000; 
Viegas, Gruenwald, \& Steigman 2000).

\begin{figure}[htb]
\begin{center}
\leavevmode
\epsfxsize=155mm
\epsfbox[78 413 566 688]{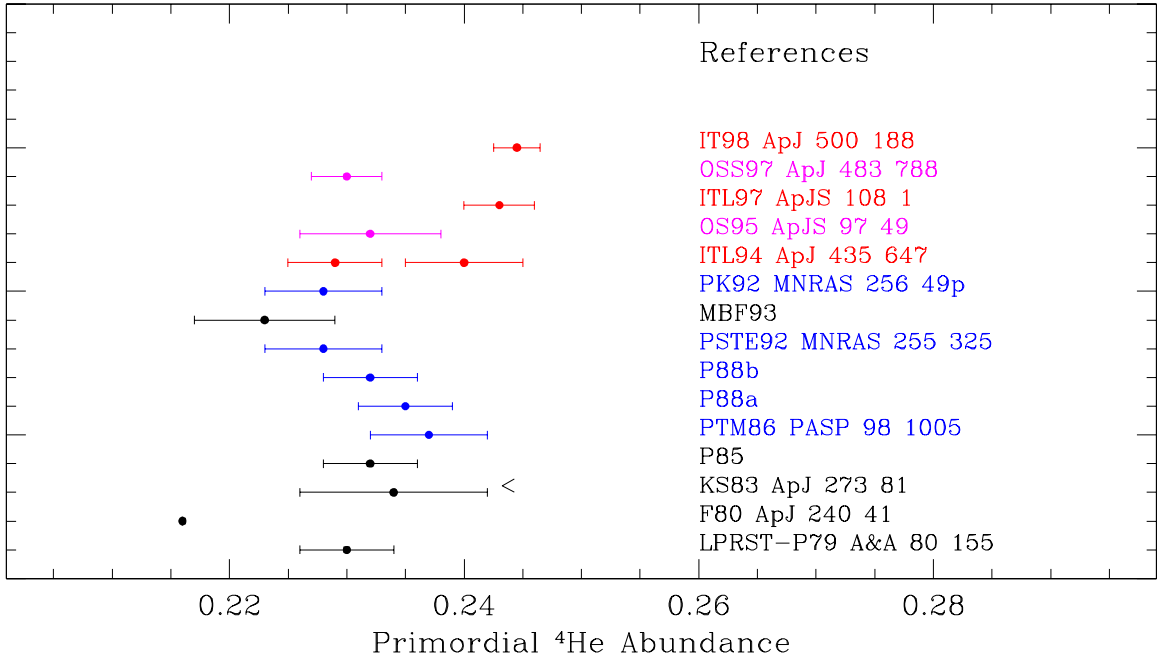}
\end{center}
\caption{Measurements of the primordial helium abundance, 1979--1998.
Horizontal points and bars give the inferred values, with references in the 
right-hand column.} 
\label{helium4_asa01}
\end{figure}

An illustration of how these factors have influenced primordial He estimates
is given in Figure~\ref{helium4_asa01}. All values are close 
to Y$_p$ = 0.24, but since the mid-1990s two streams of values have been obtained: 
a ``low'' value near 0.230, and a ``high'' value near 0.245. Moreover, the stated 
error bars are considerably smaller than this difference! The impact of these 
different values on the derived 
baryon density of the universe is discussed later.

\subsection{\bf $^3$He}
\label {3He}

$^3$He is presently almost unusable as a diagnostic of Big Bang nucleosynthesis. 
In the solar system, direct measurements
of $^3$He atoms in the solar wind and meteorites can be made, but all of this
material has passed through numerous stellar generations. 
The only other diagnostic is the 3.4~cm hyperfine transition of the $^3$He~II
ion, analogous to the 21~cm hyperfine transition 
of $^1$H~I and the 92~cm hyperfine transition of $^2$H~I,  which can be observed
in H~II regions and planetary nebulae.

To derive an abundance from the observations, an accurate model of the H~II
region or planetary nebula is needed. There is reason to doubt the suitability
of current models, because different H~II regions show a range of $^3$He
abundances that correlate with the mass of the H~II region, suggesting that the
$^3$He has been depleted locally (Rood et al. 1995). 

There are also large uncertainties in the interpretation of $^3$He abundances. 
Stars are expected to both produce and destroy $^3$He
at different stages of their life. Until very recently, the expectation was that
1--2~M$_\odot$ stars would be net producers of $^3$He while more massive stars would
destroy it, but that the overall effect would be an increase in $^3$He with 
metallicity. This was a problem, because it predicted much
more $^3$He than is observed. This {\it may} be 
resolved by recent developments in stellar evolution theory, through the
proposal by Boothroyd \& Sackmann (1999) and Sackmann \& Boothroyd (1999)
that mixing of material 
occurs deeper in stars than would be expected from standard convective
theory, and that this leads to destruction of $^3$He even in the
lower mass stars previously believed to be net $^3$He producers, 
making stars overall destroyers of this primordial isotope. 

Given these significant problems, $^3$He currently does not 
constrain primordial nucleosynthesis strongly. Bania, Rood \& Balser (2001)
describe their measurement $^3$He/H = (1.79$\pm$0.65)$\times 10^{-5}$ as a
``reasonable approximation of the primordial value,'' while Bell (2000) reached
the contrasting conclusion that ``the upper limit for the relative abundance of
primordial $^3$He ... is $^3$He$^+$/H$^+$ $\le 2\times 10^{-6}$.''

\large
\subsection{\bf $^2$H}
\label {2H} 
\normalsize

Deuterium is destroyed in stars, burning as d~+~p~$\rightarrow ^3$He~+~$\gamma$.
As there is no net source of deuterium in stars,
its abundance has decreased steadily
since the Big Bang, and any value measured today is a lower 
limit on the primordial value. Accurate
measurements of the deuterium abundance can be made in solar system bodies
and the interstellar medium of our Galaxy. The interstellar
``D$\alpha$'' line, which is displaced from the ``Lyman-$\alpha$'' line of 
$^1$H at 1216~\AA\ by $-0.33$~\AA, gives D/H~=~1.6$\times 10^{-5}$, which is
therefore a firm lower limit on the primordial value (Linsky et al. 1995).
Models of the chemical evolution of the Galaxy can be applied to estimate
how much deuterium has been destroyed over its lifetime, but
this adds more uncertainty. 

Much effort since the mid-1990's has
gone into trying to measure deuterium in the interstellar medium of
other galaxies, particularly in damped Lyman-$\alpha$ (DLA) absorption systems
superimposed on quasar spectra at high redshift.  The 
look-back times of objects at high redshifts offer the 
opportunity of looking directly into the past, and making measurements of the 
deuterium content ``back then''. Such values are expected to be closer to the
primordial one than values measured in the interstellar medium of our own
Galaxy.

Because the abundance of deuterium is so much lower than that of hydrogen, 
large column densities of hydrogen are required before deuterium can be 
detected. 
Damped Lyman-$\alpha$ clouds have high column densities
and are believed 
to be early galactic gas. Debate continues as to their nature:
are they collapsed
into flattened disks as in a mature galaxy like the Milky Way, or still
distended like the halo of the Galaxy? 
The restriction to using high column-density clouds is the source of several 
difficulties in obtaining reliable D/H measurements at high redshift.
\newline $\bullet$
Firstly, there is confusion over whether any single  absorption line is due to
deuterium or to another hydrogen line at a slightly lower redshift 
(e.g. Burles et al. 1999).
\newline $\bullet$
Secondly, although it may be possible to measure the deuterium column density
quite well, we also need to know the hydrogen
column density if we are to obtain the ratio D/H. However, since the
Lyman-$\alpha$ line is saturated (has absorbed all of the light),
it is very difficult to
measure its column density accurately. The H column density has to be
obtained from other features in the spectrum.

\begin{figure}[htb]
\begin{center}
\leavevmode
\epsfxsize=110mm
\epsfbox[68 230 566 743]{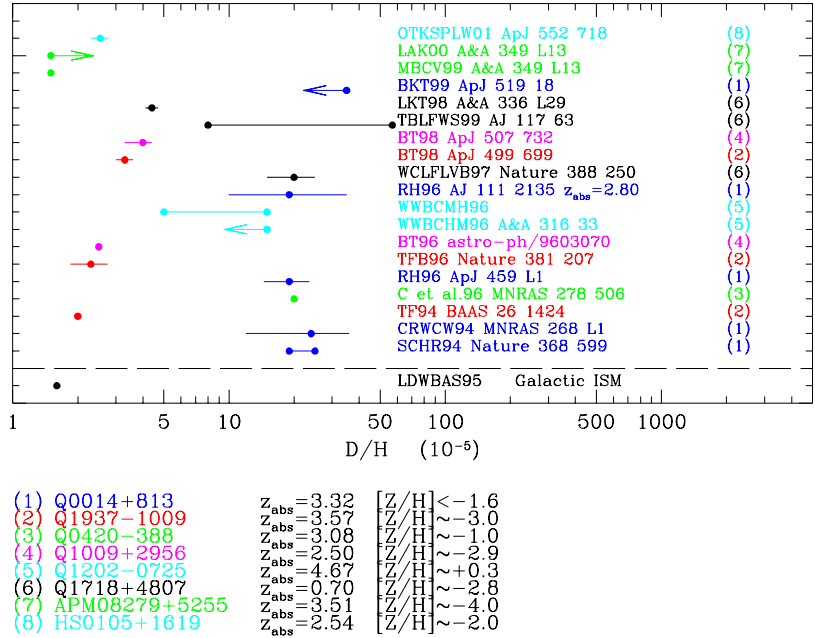}
\end{center}
\vspace{-20mm}
\caption{Measurements of the deuterium abundance in quasar absorption line
systems, 1994--1999. (See Figure~\ref{helium4_asa01} for the layout.)
The value measured in the interstellar medium (ISM) of the Milky Way is shown
at the bottom for comparison. Below the figure are recorded the quasar names,
the redshift of the measured absorption system (not the redshift of the quasar,
$z_{\rm em}$), and the metallicity Z on a logarithmic scale relative to the Sun,
where [Z/H] = log$_{10}$ (Z/Z$_{\rm Sun}$).
}
\label{deuterium_record_asa01}
\end{figure}

Figure~\ref{deuterium_record_asa01} shows measurements towards a
range of quasars over 1994-1999. 
A considerable range of values has been obtained, including
``low'' values around D/H = (2--5)$\times 10^{-5}$, and 
``high'' values around D/N = (15--30)$\times 10^{-5}$.

The high and low values have very different implications. As any real 
measurement of deuterium sets a lower limit on the primordial abundance, if 
{\it any} of the ``high'' measurements is correct, then the primordial value must be (at least!)
an order of magnitude higher than that inferred from the Galactic ISM.
However, the difficulty of knowing whether absorption just
shortward of the damped Lyman-$\alpha$ line is really due to deuterium
weakens the arguments for very high D/H ratios.  ``Low'' values
have been measured in several systems, in particular the ones 
with low metallicity [Z/H] $\sim$ $-$3.0, $-$2.9, and $-$2.8. 
For the ``high'' values to be correct, deuterium would have to be destroyed 
very {\it quickly} in the low metallicity systems despite their low 
metallicities indicating very {\it little} stellar processing. 
This seems unlikely (Tosi et al. 1998; Prantzos \& Ishimaru 2001). 
It is more likely that the lowest
metallicity systems are more reliable measurements, and that the ``high''
values are misleading due to the presence of additional $^1$H absorption lines.
Data are still being taken; it is perhaps a matter of opinion whether
the issue has been settled, but it is probable
that the ``low'' deuterium measurements will win out over the ``high'' ones once
the dust settles. (See references to Figure~\ref{deuterium_record_asa01}.)

\large
\subsection{\bf $^7$Li}
\label {7Li}
\normalsize

Lithium is measurable in the spectra of main-sequence halo dwarfs.
$^7$Li is much less fragile than $^2$H, and as the outer layers
of solar-mass metal-poor dwarfs do not mix to great depths, the surface
Li in these stars is widely believed to be little changed since they
formed $\sim$13~Gyr ago. The main evidence in support of this
is empirical: most solar-temperature, metal-poor dwarfs have 
almost the same Li abundance despite having different amounts
of other elements and different masses (e.g. Spite \& Spite 1982).
It is unlikely that all such stars could deplete or
supplement their Big Bang Li complement and {\it still} end up with the same
value after $\simeq$14~Gyrs. 
Theoretical models of stars differ in the amount of Li they preserve, and
it is just conceivable that some mechanisms of Li destruction might conspire
to deplete all stars by the same amount. However, the observations can
tolerate little Li depletion even over the long lifetimes of these stars. 
For processing by rotationally-induced mixing, Ryan, Norris, \& Beers (1999) 
and Pinsonneault et al. (2001) infer limits $<0.1$ and $\simeq$0.2~dex 
respectively. The difference in these values
depends on (A) the frequency of the few stars with Li abundances significantly 
lower than the rest, (B) the size distribution of their deficiencies, and
(C) whether the deviations are in fact due to the proposed mechanism 
(rotationally-induced mixing) and not some other.
Small amounts of Li {\it production} may also be present in the stars 
that formed more recently from nucleosynthetically enriched material. This
can be tracked analogously to the way He production is tracked by d$Y$/d$Z$. 
The main
uncertainty remaining for Li is how accurately we know the temperatures of the
stellar atmospheres, and therefore how well we can compute the Li abundances
from their spectra.
The estimate of the primordial Li abundance by Ryan et al. (2000) is 
$n$(Li)/$n$(H) = 1.23$^{+0.68}_{-0.32} \times 10^{-10}$.

Note, however, that even though the primordial Li abundance may be reasonably
well known, there are several significant deficiencies in our understanding of 
Li processing in Pop~II and Pop~I stars. About 7\% of Pop~II stars which would
be expected to have normal Li abundances show extensive (essentially complete) 
destruction of the element.  Investigations of other elements in these stars 
(Norris et al. 1997b; Ryan, Norris \& Beers 1998) show heterogeneous behaviour
and indicate that no single characteristic besides Li deficiency unites these
stars. Ryan et al. (2001c) have suggested they may be 
blue-stragglers-to-be,  sub-turnoff-mass products of classical blue-straggler
formation mechanisms.
There is also an unexplained contrast between the highly uniform Li abundances 
measured in halo field stars (Ryan et al. 1999) 
and the large spread of measurements for subgiants in
the halo globular cluster M92 (Boesgaard et al. 1998).

\large
\subsection{\bf Putting it all together}
\label {Putting it all together}
\normalsize

We now put the observational constraints together and 
examine the concordance with theory. The standard
Big Bang model assumes a uniform density and predicts the yields
of the primordial isotopes for different values of that density.\footnote{The
density may be expressed variously as the baryon-to-photon ratio, $\eta$, or the
baryon density as a fraction of the critical density, $\Omega_B$, where
$\Omega_B = 0.014 \times {{\eta}\over{10^{-10}}} \times 
({{50~{\rm km~s^{-1}~Mpc^{-1}}}\over {H_0}})^2$.}
The predicted yield of $^4$He is close to 24\% for a wide range of densities,
and the fact that this is so close to the real universe is a remarkable
accomplishment of the model. It is even more remarkable that for $^2$H and 
$^7$Li the abundances are also in accord. 
For three isotopes spanning a range
of 1 billion in abundance, the model is broadly in accord with observations.

\begin{figure}[ht]
\begin{center}
\leavevmode
\epsfxsize=070mm
\epsfbox[255 166 567 707]{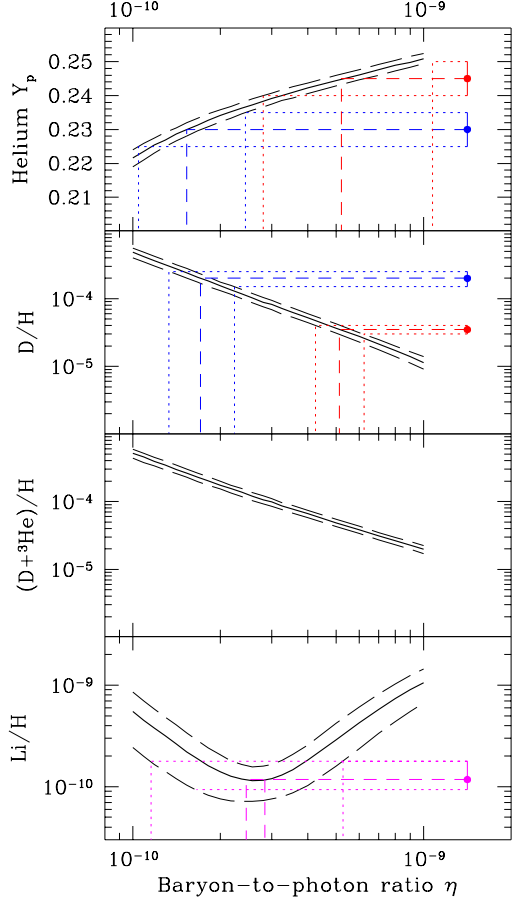}
\end{center}
\caption{{\it Solid curves}: Calculations of yields of Big Bang nucleosynthesis
for $^4$He, $^2$H, the sum of $^2$H and $^3$He, and $^7$Li, for different 
assumed baryon-to-photon ratios. {\it Long dash curves}: 95\% confidence range
for the nucleosynthesis calculations, primarily reflecting uncertainties in the
nuclear reaction rates. {\it Data symbols}: current observational values,
split for $^4$He and $^2$D into ``high'' and ``low'' values to reflect current
systematic uncertainties.
}
\label{bbn_abunds}
\end{figure}

Figure~\ref{bbn_abunds} presents the situation in more detail, and illustrates the
different values of the derived $^4$He and $^2$H abundances, and the $^7$Li
value. No $^3$He observations are shown.
The predicted yield of Li is not monotonic with density, as a second
nucleosynthesis pathway becomes effective for $\eta > 2\times 10^{-10}$. The 
well-defined Li abundance is therefore consistent with a wide range of 
density values. Two observational results are given for 
each of $^4$He and $^2$H, the ``high'' and ``low'' values discussed above. 
Corrections for systematic errors in the 
$^4$He analysis favour ``high'' values, and the study of more damped
Lyman-$\alpha$ systems favour the ``lower'' range for D, 
giving a value for $\eta \simeq 5\times 10^{-10}$. 
(The slightly lower 95\% confidence interval given by Ryan et al. (2000),
$\eta = (1.7-3.9)\times 10^{-10}$,
was based on a primordial $^4$He abundance intermediate between the two observational
values shown here.)

An independent measurement of the universal baryon density is possible from the
power spectrum of the cosmic microwave background radiation (CMBR).
The higher-order fluctuations above the main ($\simeq 1^\circ$) peak are influenced by the
propagation of sound waves, which reflects the baryon density. Measurements published in
2000 giving $\eta = 10\times 10^{-10}$ raised the spectre of incompatible  
Big Bang nucleosynthesis (BBN) and CMBR measurements of the baryon density (Hogan 2000).
Fortunately, the earlier values have been superseded with the addition of more data
such that the CMBR measurements now suggest a 95\% confidence interval
$\eta = (3.4-7.9)\times 10^{-10}$, which is certainly compatible with the
BBN values cited above (Netterfield et al. 2001).

For an assumed Hubble constant of 
63~km~s$^{-1}$~Mpc$^{-1}$, the value $\eta = 5\times 10^{-10}$ corresponds to
$\Omega_B = 0.043$. Not only is this well below the closure
density, it is also a factor of ten below the matter density 
$\Omega_M \simeq 0.35$ inferred from other cosmological studies (Efstathiou et al. 1999),
pointing to the continuing need for non-baryonic dark matter.

\clearpage

\section*{Acknowledgments}

It is a pleasure to thank the organising committee of the 2001 Conference
of the Astronomical Society of Australia for their support of my attendance
at the meeting. The suggestions of an anonymous referee were 
gratefully received.

\section*{References}


\reference Allende Prieto, C., et al. 2000, AJ, 120, 1516
\reference Aoki, W., Norris, J. E., Ryan, S. G., Beers, T. C., \& Ando, H. 2002, subm.
\reference Ballantyne, D.\ R., Ferland, G.\ J., \& Martin, P.\ G.\ 2000, ApJ, 536, 773
\reference Bell, M.\ B.\ 2000, ApJ, 531, 820
\reference Bania, T.\ M., Rood, R.\ T., \& Balser, D.\ S.\ 2001, Bull. AAS, 198, 4106
\reference Barbuy, B., Cayrel, R., Spite, M., Beers, T.C., Spite, F., Nordstrom, B., \& Nissen, P. E. 1997, A\&A, 317, L63
\reference Beers, T. C., Preston, G. W. \& Shectman, S. A. 1985,  AJ,  { 90}, 2089
\reference Beers, T. C., Preston, G.  W., \& Shectman, S.  A. 1992,  AJ, 103, 1987
\reference Bessell, M. S. \& Norris, J. E. 1984, ApJ,  { 285}, 622
\reference Boesgaard, A. M., Deliyannis, C. P., Stephens, A., \& King, J. R. 1998, ApJ, 493, 206
\reference Boothroyd, A. I. \& Sackmann, I.-J. 1999, ApJ, 510, 232
\reference Brown, J. A., \& Wallerstein, G. 1989, AJ,  { 98}, 1643
\reference Burles, S., Kirkman, D., \& Tytler, D. 1999, ApJ, 519, 18
\reference Burles, S. \& Tytler, D. 1996, astro-ph/9603070
\reference Burles, S. \& Tytler, D. 1998, ApJ, 499, 699
\reference Burles, S. \& Tytler, D. 1998, ApJ, 507, 732
\reference Carswell, R. F., Rauch, M., Weymann, R. J., Cooke, A. J., \& Webb, J. K. 1994, MNRAS, 268, L1
\reference Carswell, R. F. et al. 1996, MNRAS, 278, 506
\reference Cowan, J. J., Burris, D. L., Sneden, C., McWilliam, A., \& Preston, G. W. 1995, ApJ, 439, L51
\reference Edmunds, M. G., \& Phillipps, S. 1997, MNRAS, 292, 733
\reference Efstathiou, G., Bridle, S. L., Lasenby, A. N., Hobson, M. P., \& Ellis, R. S. 1999, MNRAS, 303, L47
\reference French, H. B. 1980, ApJ, 240, 41
\reference Hartwick, F. D. A. 1976, ApJ,  209, 418
\reference Hauge, O. 1972, Sol.Phys., 27, 286
\reference Hogan, C. 2000, Nature, 408, 47
\reference Izotov, Y. I., \& Thuan, T. X. 1998, ApJ, 500, 188
\reference Izotov, Y. I., Thuan, T. X. \& Lipovetsky, V. A. 1994, ApJ, 435, 647
\reference Izotov, Y. I., Thuan, T. X. \& Lipovetsky, V. A. 1997, ApJS, 108, 1
\reference Kunth, D. \& Sargent, W. L. W 1983, ApJ,  273,  81
\reference Mathews, G. J., and Cowan, J. J. 1990, Nature, 345, 491
\reference Lang, K. R. 1980, Astrophysical formulae, (Springer-Verlag, Berlin)
\reference Lequeux, J., Rayo, J. F., Serrano, A., Peimbert, M., \& Torres-Peimbert, S., 1979, A\&A, 80, 155
\reference Levshakov, S.\ A., Agafonova, I.\ I., \& Kegel, W.\ H.\ 2000, A\&A, 355, L1
\reference Levshakov, S. A., Kegel, W. H., \& Takahara, F. 1998, A\&A, 336, L29
\reference Linsky, J. L., Diplas, A., Wood, B. E., Brown, A., Ayres, T. R., \& Savage, B. D. 1995, ApJ, 451, 335
\reference Magain, P. 1995, A\&A, 297, 686
\reference Mathews, G. J., Boyd, R. N., \&\ Fuller, G. M. 1993, ApJ, 403, 65
\reference Mathews, G. J., and Cowan, J. J. 1990, Nature, 345, 491
\reference Molaro, P., Bonifacio, P., Centurion, M. \& Vladilo, G. 1999, A\&A, 349, L13
\reference Nakamura, F. \& Umemura, 1999, ApJ, 515, 239
\reference Netterfield et al. 2001, astro-ph/0104460
\reference Norris, J. E. 1999, in The Third Stromlo Symposium: The Galactic Halo, 
ASP Conf Ser. 165, 213
\reference Norris, J. E., Ryan, S. G., \& Beers, T. C. 1997a, ApJ, 488, 350
\reference Norris, J. E., Ryan, S. G., \& Beers, T. C. 1997c, ApJ, 489, L169
\reference Norris, J. E., Ryan, S. G., \& Beers, T. C. 1998, ApJS, 123, 639
\reference Norris, J. E., Ryan, S. G., \& Beers, T. C. 2001, ApJ, 562, in press
\reference Norris, J. E., Ryan, S. G., Beers, T. C., \& Deliyannis, C. P. 1997b, ApJ, 485, 370
\reference Olive, K. A. \& Skillman, E. D. 2001, New Astronomy, 6, 119
\reference Olive, K. A., \& Steigman, G. 1995, ApJS, 97, 49
\reference Olive, K. A., Steigman, G. \& Skillman, E. D. 1997, ApJ, 483, 788
\reference O'Meara, J.\ M., Tytler, D., Kirkman, D., Suzuki, N., Prochaska, J.\ X., Lubin, D., \& Wolfe, A.\ M.\ 2001, ApJ, 552, 718
\reference Pagel, B. E. J. 1988a, Astro-Particle Physics, A. de Rujula, D. Nanapoulos, \& P. A. Shaver (eds) (Singapore: World Scientific), 399
\reference Pagel, B. E. J. 1988b, in Evolutionary phenomena in galaxies,  J. Beckman \& B. E. J. Pagel (eds) (Cambridge: Cambridge Uni. Press), 201
\reference Pagel, B. E. J., and Kazauskas, A. 1992, MNRAS, 256, 49p
\reference Pagel, B. E. J., Simonson, E. A., Terlevich, R. J., and Edmunds, M. G. 1992, MNRAS, 255, 325
\reference Pagel, B. E. J., \& Tautvai\v{s}ien\.{e}, G. 1997, MNRAS, 288, 108
\reference Pagel, B. E. J., Terlevich, R. J., \& Melnick, J. 1986, PASP, 98, 1005
\reference Peimbert, M. 1985, in Star Forming Dwarf Galaxies, D. Kunth, T. X. Thuan \& J. T. T. Van (eds) (Paris: Ed. Fornti\`eres, 403
\reference Pinsonneault, M. H., Walker, T. P., Steigman, G., \& Narayanan, V. K. 2001, astro-ph/0105439
\reference Prantzos, N. 1994, A\&A, 284, 477
\reference Prantzos, N. \& Ishimaru, Y. 2001, A\&A, 376, 751
\reference Rood, R. T., Bania, T. M., Wilson, T. L., \& Balser, D. S. 1995, in The Light Element Abundances, ed. P. Crane (Berlin: Springer), 201
\reference Rossi, S., Beers, T. C., \& Sneden, C. 1999, ASP Conf. Ser. 165, 1999
\reference Rugers, M., \& Hogan, C. J. 1996a, ApJ, 459, L1
\reference Rugers, M., \& Hogan, C. J. 1996b, AJ, 111, 2135
\reference Ryan, S. G.  2001 ``The Influence of Binaries on Stellar Population Studies''
D. Vanbeveren (ed), (Kluwer, Dordrecht), 491
\reference Ryan, S. G., Aoki, W., Blake, L. A. J., Norris, J. E., Beers, T. C., Gallino, R., Busso, M. \& Ando, H. 2001b, Mem.S.A.It., in press
\reference Ryan, S. G., Aoki, W., Norris, J. E., Beers, T. C., Gallino, R., Busso, M. \& Ando, H. 2001a, Nucl.Phys.A, 688, 209c
\reference Ryan, S. G., Beers, T. C., Kajino, T., \& Rosolankova, K. 2001c, ApJ, 547, 231
\reference Ryan, S. G., Beers, T. C., Olive, K. A., Fields, B. D., \& Norris, J. E. 2000, ApJ,  530, L57
\reference Ryan, S. G. \& Norris, J. E. 1991a, AJ, 101, 1865
\reference Ryan, S. G. \& Norris, J. E. 1991b, AJ, 101, 1835
\reference Ryan, S. G., Norris, J. E., \& Beers, T. C. 1998, ApJ, 506, 892
\reference Ryan, S. G., Norris, J. E., \& Beers, T. C. 1999, ApJ, 523, 654
\reference Ryan, S. G., Norris, J. E. \& Bessell, M. S. 1991, AJ, { 102}, 303
\reference Sackmann, I.-J., and Boothroyd, A. I. 1999, ApJ, 510, 217
\reference Shetrone, M. D. 1996, AJ, 112, 2639
\reference Smith, V. V., Lambert, D. L., \& Nissen, P.-E. 1993, ApJ, 408, 262
\reference Songaila, A., Cowie, L. L., Hogan, C. J., \& Rugers, M. 1994, Nature, 368, 599
\reference Spite, F. \& Spite, M. 1982, A\&A,  115, 357
\reference Spite, M., \& Spite, F. 1978, A\&A, 67, 23
\reference Timmes, F. X., Woosley, S. E., \& Weaver, T. A. 1995, ApJS, 98, 617
\reference Tinsley, B. M. 1979, ApJ, 229, 1046
\reference Tosi, M., Steigman, G., Matteucci, F. \& Chiappini, C. 1998, ApJ, 498, 226
\reference Travaglio, C., Galli, D., Gallino, R., Busso, M., Ferrini, F., \& Straniero, O. 1999, ApJ, 521, 691
\reference Truran, J. W. 1981, A\&A, 97, 391
\reference Tumlinson, J. \& Shull, J. M. 2000, ApJ, 528, L65
\reference Tytler, D., Burles, S,. Lu, L,. Fan, X.-M., Wolfe, A., \& Savage, B. D. 1999, AJ, 117, 63
\reference Tytler, D. \& Fan, X.-M. 1994, BAAS, 26, 1424
\reference Tytler, D., Fan, X.-M., \& Burles, S. 1996, Nature, 381, 207
\reference Viegas, S.\ M., Gruenwald, R., \& Steigman, G.\ 2000, ApJ, 531, 813
\reference Wampler, E. J., Williger, G. M., Baldwin, J. A., Carswell, R. F., Hazard, C. \& McMahon, R. G. 1997, A\&A, 316, 33
\reference Webb, J. K., Carswell, R. F., Lanzetta, K. M., Ferlet, R., Lemoine, M., Vidal-Madjar, A., \& Bowen, D. V. 1997, Nature, 388, 250
\reference Woosley, S. E. \& Weaver, T. A. 1995, ApJS, 101, 181
\reference Yoshii, Y. 1981, A\&A, { 97}, 280



\end{document}